\documentclass{emulateapj}

\def\apm{APM~08279+5255}
\def\sapm{APM08}
\def\asec{$\arcsec$}
\def\fasec{$\farcs$}
\def\kms{km s$^{-1}$}

\def\asec{$\arcsec$}

\shorttitle{High Angular Resolution SMA Imaging of \apm}
\shortauthors{Krips et al.}

\begin{document}

\title{SMA High Angular Resolution Imaging of the Lensed Quasar \apm\,}

\author{M.~Krips\altaffilmark{1,2}, A.B.~Peck\altaffilmark{1},
K.~Sakamoto\altaffilmark{3}, G.B.~Petitpas\altaffilmark{1},
D.J.~Wilner\altaffilmark{2}, S.~Matsushita\altaffilmark{4} and
D.~Iono\altaffilmark{3}}

\altaffiltext{1}{Harvard-Smithsonian Center for Astrophysics, SMA
project, North A'ohoku Place, Hilo, HI, 96720; mkrips@cfa.harvard.edu}

\altaffiltext{2}{Harvard-Smithsonian Center for Astrophysics, 60
Garden Street, Cambridge, MA 02138}

\altaffiltext{3}{National Astronomical Observatory of Japan, Mitaka,
Tokyo, 181-8588, Japan}
%\altaffiltext{4}{Institut Radio Astronomie Millimetrique (IRAM), 300
%rue de la Piscine, Domaine Universitaire, 38406 Saint Martin d'H\`eres,
%France } 
\altaffiltext{4}{Academia Sinica Institute of Astronomy and
Astrophysics, P.O. Box 23-141, Taipei 106, Taiwan, R.O.C.}

\begin{abstract}
We present Submillimeter Array observations of the z=3.91
gravitationally lensed broad absorption line quasar \apm\, which
spatially resolve the 1.0~mm (200~$\mu$m rest-frame) dust continuum
emission. At 0\fasec4 resolution, the emission is separated into two
components, a stronger, extended one to the northeast ($46\pm5$~mJy)
and a weaker, compact one to the southwest ($15\pm2$~mJy). We have
carried out simulations of the gravitational lensing effect
responsible for the two submm components in order to constrain the
intrinsic size of the submm continuum emission. Using an elliptical
lens potential, the best fit lensing model yields an intrinsic
(projected) diameter of $\sim$80~pc, which is not as compact as the
optical/near-infrared (NIR) emission and agrees with previous size
estimates of the gas and dust emission in \apm. Based on our estimate,
we favor a scenario in which the 200~$\mu$m (rest-frame) emission
originates from a warm dust component (T$_{\rm d}$=150-220~K) that is
mainly heated by the AGN rather than by a starburst (SB). The flux is
boosted by a factor of $\sim$90 in our model, consistent with recent
estimates for \apm.

\end{abstract}


\section{Introduction}

\apm\, (=APM08) is a strongly lensed broad absorption line (BAL)
quasar \citep{irwin98,lewis98} with a very powerful active galactic
nucleus \citep[AGN;][]{soifer87}.  This combination makes it an
extremely bright object despite its redshift of $z$=3.91
\citep{downes99}.  Its bolometric luminosity is thought to be
$\sim$5$\times$10$^{13}$~L$_{\odot}$, amplified by up to a factor of
100 by a foreground galaxy which has yet to be identified.  Due to the
large amplification, broad absorption lines have been detected even in
the X-ray \citep{chartas02}. Ground-based, {\it Chandra} and HST
observations have suggested that \sapm\, consists of at least two
components separated by $\sim$$0\farcs4$. \citet[=I99]{ibata99} and
\citet[=E00]{egami00} later detected a third image (their image C),
and argued that it is likely a third lensed image of the background
QSO instead of the lensing galaxy.  \cite{lewis02b} obtained optical
spectra of the three images using HST/STIS, and showed that the three
spectra are quite similar.  This indicates that the image C is indeed
a third lensed image of \sapm. Due to the magnification by the lens,
several molecular lines have been detected in \sapm\, \citep[e.g.,][]
{weiss07,guelin07,riecher06,wagg06,garcia06,wagg05,lewis02a,downes99},
suggesting a molecular gas mass of M$_{\rm
gas}$(H$_2$)=8$\times$10$^{10}$~$m^{-1}$~M$_\odot$
\citep[][$m$$\equiv$magnification factor]{riecher06}.  The dust mass
is estimated to be M$_{\rm dust}$=5$\times$10$^8$~$m^{-1}$~M$_\odot$
\citep[][W07]{downes99,weiss07}. However, despite the increasing
amount of mm-data, no tight constraints have been set on the size of
the dust and/or gas emission which may help to characterize the
heating source of the dust and gas, i.e., whether it is in the form of
an AGN and/or starburst (SB). Only recently have W07 presented some
indirect evidence through brightness temperature arguments that the
molecular line and dust emission originate from a compact region with
a radius of 100-200~pc. In this {\it Letter}, we present observations
from the Submillimeter Array (SMA) with sufficient resolution
(0\fasec4) to resolve the 1.0~mm dust continuum emission in \sapm, and
lensing models that constrain its intrinsic source size.

\section{Observations}
We observed the 1.0~mm (200~$\mu$m rest-frame) continuum emission in
\sapm\, with the SMA using seven of the eight 6~m antennas in two
configurations, with baselines ranging from 20 to 500 meters, in
November and December 2006. This results in a synthesized beam of
0\fasec42$\times$0\fasec35 at a position angle (PA) of 141$^\circ$
using robust weighting. The LO frequency was 302.4~GHz (1.0~mm), with
each sideband centered 5~GHz to either side. This frequency setup was
chosen so that the [\ion{N}{2}] line (1461~GHz rest frame) falls
within the LSB. However, no significant singal of this line has been
found above a 3$\sigma$ threshold of 9~Jy~beam$^{-1}$~\kms, integrated
over a velocity range of 1000~\kms\ around the expected [\ion{N}{2}]
frequency. We used Titan as a flux calibrator, 3C111, 3C454.3,
J0927+390 and 3C279 as bandpass calibrators, and J0927+390 as a gain
calibrator. J0753+538 has been frequently observed throughout the
night as a test source to assess the quality of the gain
calibration. The phase rms on the gain calibrator is $\leq$40~degrees,
resulting in a positional uncertainty of $<$0\fasec1 in the test map
of J0753+538, which is similar to theoretical noise considerations
\citep{reid88}, and to a 'seeing' of $<$0.03$''$ ($\equiv$difference
between the apparent size of the test quasar and the beam size). The
accuracy of the flux calibration is around 20\%.  Weather conditions
were good with an atmospheric opacity of $\tau$(225~GHz)=0.06-0.12
(recorded at the nearby Caltech Submillimeter Observatory).  The data
were written into the MIR format originally developed at the Owens
Valley Radio Observatory \citep{scoville93}, and reduced using
SMA-specific MIR tasks in IDL.  Following calibration, the data were
exported to FITS and imaging and further analysis were done using
GILDAS.  Both 2~GHz sidebands were merged together in the uv-plane to
image the 1.0~mm continuum emission, resulting in an effective rms
noise of 1.7~mJy~beam$^{-1}$.

\section{Results}
\begin{figure}[!]
\centering
%\rotatebox{-90}{\resizebox{4.0cm}{!}{\includegraphics{fig1a.eps}}}
\rotatebox{-90}{\resizebox{3.8cm}{!}{\includegraphics{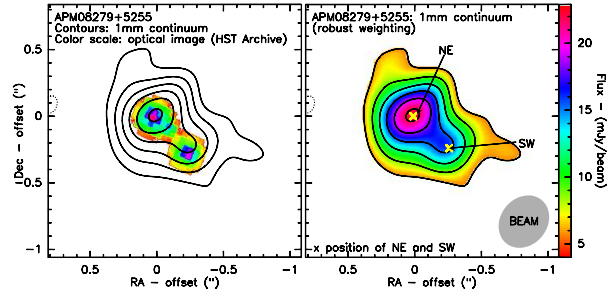}}}
%\hspace*{0.2cm}
%\rotatebox{-90}{\resizebox{4.0cm}{!}{\includegraphics{fig1b.eps}}}
%\rotatebox{-90}{\resizebox{6.0cm}{!}{\includegraphics{fig1b.eps}}}
\caption{{\it Left:} HST image of \sapm\, observed with the
WFPC2/F814W filter (STScI archive). The black contours represent the
1mm continuum emission. {\it Right:} Continuum image at the observed
wavelength of 1.0~mm of \sapm\ with robust weighting. Contours start
at $\pm$3$\sigma$ in steps of 2$\sigma$; 1$\sigma$ corresponds to
1.7~mJy~beam$^{-1}$. The yellow crosses mark the (deconvolved)
position of the two lensed images; they are thus slightly more
separated than the peaks of the contour image. The image is referenced
to component NE (Table~\ref{tab1}).}
\label{fig1}
\end{figure}

The 1.0~mm continuum in \sapm\, is clearly detected at high angular
resolution ($\sim$0\fasec4) and high sensitivity (SNR$\geq$9;
Fig.~\ref{fig1}). Using robust weighting, the two lensed images,
labeled NE (north-eastern) and SW (south-western) respectively
(Fig~\ref{fig1}), are clearly separated (by $\sim$0\fasec4) from each
other. The image positions listed in Table~\ref{tab1} are measured
after the image was deconvolved with the synthesized beam. Both
absolute positions and the separation are consistent with previous
results within the uncertainties (I99, E00). A comparison of the
robustly and naturally weighted (angular resolution $\sim$0\farcs6)
maps indicates similar total fluxes of $\sim$60$\pm$12~mJy
(Table~\ref{tab1}). Taking the (single dish) fluxes of 24~mJy at
1.35~mm and of 75~mJy at 0.850~mm and assuming a spectral index of
2.46 (\cite{lewis98}), a flux of 50~mJy is expected at 1~mm which is,
within the uncertainties, in agreement with the SMA value. This
indicates that no flux has been resolved out in our data and the dust
emission at 200~$\mu$m rest-frame cannot be very extended, similar to
recent results in \sapm\, by W07.

The deconvolved size (=FWHM) of the emission in NE appears to be
slightly more extended (by $\sim$30\%) than the synthesized beam
(Table~\ref{tab1}). An extension in NE is further supported in
comparing the peak flux with the spatially integrated flux
(Table~\ref{tab1}). While they differ by a factor of $\sim$2 for NE,
they are very similar in SW. This suggests that SW is still unresolved
at the high angular resolution of the robustly weighted map. We derive
a peak flux ratio of (NE/SW)$_{\rm peak}\approx$1.4 and a spatially
integrated flux ratio of (NE/SW)$_{\rm int}\approx$3. The latter is
significantly higher than the value found in the (compact) optical/NIR
emission of $\sim$1.2 (e.g., I99,E00), indicating differential lensing
effects. The morphology of the 200~$\mu m$ rest-frame dust emission is
in good agreement with that of the CO(1--0) emission of \sapm\,
recently published by \cite{walter06}. Based on their CO(1--0) map
\cite[Fig.~4 in][]{walter06}, the NE/SW peak flux ratio is estimated
to be $\sim$1.5 and the FWHM of NE is estimated to be
0\fasec3-0\fasec4, which is slightly larger than their beamsize, while
SW is clearly unresolved.

\section{Simulations of the lensing effect}
\label{ep3}
Previous lensing models for \sapm, obtained by fitting the (compact)
NIR emission in E00, fail to reproduce the (extended) emission seen in
our submm observations, as in their extended emission models the
southern image turns out to be too strong and too extended. Therefore,
we have run new simulations\footnote{Assumed cosmology: $\Omega_{\rm
M}$=0.3, $\Omega_{\rm R}$=0.0 and $\Omega_{\rm V}$=0.7 for a Hubble
constant of H$_0$=71\kms~Mpc$^{-1}$ \citep{spergel03}, giving an
angular-size distance of $D_A$=1.47~Gpc and 1$''$=7~kpc.} of the
gravitational lensing effect using the code developed by Krips \& Neri
\citep[][available in GILDAS]{krips05}. We concentrate our simulations
on an elliptical lens potential ($\equiv$EP) similar to previous
studies (e.g., E00), although we also tried a single isothermal sphere
(SIS) as a lens potential. Although the SIS model fits our submm data,
we discarded the SIS models based on the assumption that \sapm\, is a
three-image system and because of the inability of the SIS model to
produce an odd number of lensed images.

We varied the parameters of the lensing potential as well as the size
of the unlensed source in a reasonable parameter space (see
Table~\ref{tab2}) until the best agreement between the observed and
simulated maps was found, i.e., a model with the lowest reduced
$\chi^2$ test ($\leq$1). We used the position of the two lensed images
(see Table~\ref{tab1}), their peak flux ratio (NE/SW=1.4$\pm$0.2) as
well as their (lensed) shape (FWHM$\simeq$0\fasec5$\pm$0\fasec1 in NE
and FWHM$\leq$0\fasec2 in SW) as constraints. We augmented this set of
constraints with those from the (compact) optical/NIR emission (e.g.,
I99, E00; Fig.~\ref{fig1}), accounting also for a third lensed image
close to NE. This results in 9 observational constraints versus 8 free
lensing parameters. We processed the simulated lensed images through a
spatial filter defined by the uv-coverage of the observations.  This
is a more robust approach than simply convolving the simulated data
with the synthesized beam of the observations as it also accounts for
possible resolution effects \citep[see also][]{krips05}. However,
based on this set of observational constraints, we could not find any
lensing solution that reproduces the optical/NIR and our submm
observations. We doubt that this inconsistency is due to mismatched
angular resolutions between the optical/NIR and our observations.
When starting with an elliptical potential that yields three lensed
images and fits the optical/NIR data, NE always turns out to be either
too strong or not extended enough compared to SW to match the submm
data, or, equivalently, SW is either too extended or not strong enough
compared to NE to be consistent with the submm emission. Also, models
that fit the submm emission always make the lensed optical/NIR image
NE too strong or too extended compared to SW.

\begin{figure*}[!]
\centering
\rotatebox{-90}{\resizebox{!}{\hsize}{\includegraphics{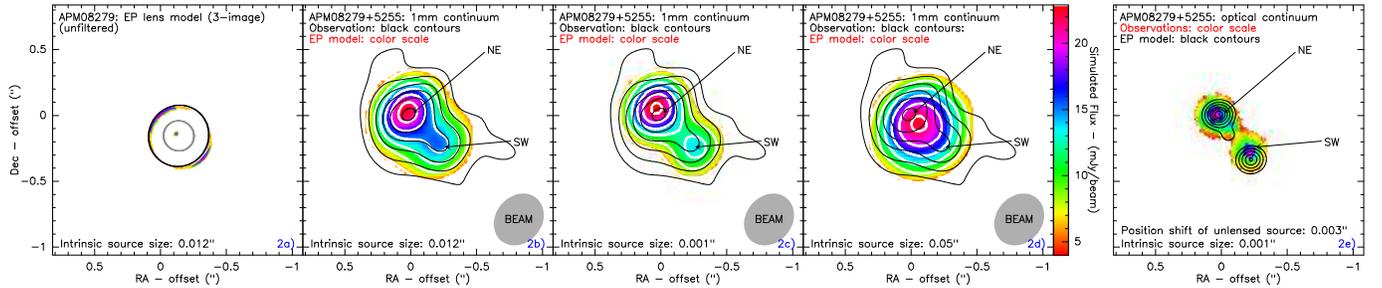}}}
\caption{Simulations ({\it color scale and white contours}) based on
an elliptical lens model with three lensed images: EP(3).  The
simulated emission is overlaid with the observed emission ({\it black
contours}). Contours are the same as in Fig.~\ref{fig1}. The model
with the best-fit source size for the submm data are shown in {\it
2a)} (unfiltered) and in {\it 2b)} (spatially filtered). {\it 2c)} and
{\it 2d)} columns show the best-fit model with the smallest and
largest unlensed source sizes, while {\it 2e)} shows the model for the
optical/NIR case (smoothed to 0\fasec15). The models in {\it 2c)} and
{\it 2d)} refers to the (unlensed) submm position; the (unlensed)
optical/NIR position is offset by $\sim0\farcs003$ in {\it 2e)} from
the (unlensed) submm position (see Section~\ref{ep3}).}
\label{fig2}
\end{figure*}

A solution to reproduce the optical/NIR and submm data with the same
lensing model is to allow different positions between the unlensed
optical/NIR and submm sources of emission. This actually increases the
number of free lensing parameters to 10. However, we have to handle
the observed positions of the lensed images in the optical/NIR and
submm case as independent parameters as well, so that the
observational constraints also increase to 11. Even though the
positional accuracy of the submm and the optical/NIR observations do
not yet allow us to substantiate the positional offset
($\sim$0\fasec08$\pm$0\fasec07) between the (lensed) optical/NIR and
submm emission and thus also between the unlensed positions, this is
still a reasonable approach since it yields a lens model that fits
both optical/NIR and submm data. This lens model, called EP(3), is
plotted in Fig.~\ref{fig2}a and \ref{fig2}b and listed in
Table~\ref{tab2}. For comparison, Fig.~\ref{fig2}c and \ref{fig2}d
respectively show simulations in which the unlensed submm source is
smaller and larger than the best-fit size. The lens parameters and the
position of the unlensed submm source are the same as in
Fig~\ref{fig2}a and \ref{fig2}b.  As the optical/NIR source is very
compact in its intrinsic size, the simulation in Fig.~\ref{fig2}c is
the one to be compared with the optical/NIR observations. The NE/SW
ratio for this compact case, however, disagrees with the optical/NIR
observations, and the peak separation is clearly wider than what is
observed. To correctly reproduce the optical/NIR data, the position of
the unlensed source has to be shifted by $\sim$0.003$''$
($\equiv$21~pc; Fig.~\ref{fig2}e). Such an offset between optical/NIR
and submm emission could be a consequence of asymmetrically
distributed submm emission in \sapm. The (best-fit) intrinsic source
diameter of the submm emission is $\sim$0.012$"$ ($\equiv$84~pc) and
the magnification factor $\sim$90, consistent with recent results from
W07.

\section{Nature of the dust emission}
W07 and B06 \citep{beelen06} both present a two temperature model for
the dust emission in \sapm\, with an extended 'cold' (T$_d$=50-70~K)
and a compact 'warm' (T$_d$=160-220~K) component.  The warm dust
component is assumed to be heated by the AGN up to a maximum radius of
60-130~pc (W07). W07 also estimate that the dust heating from the AGN
can still reach $\sim$65~K at a radius of 350~pc and may dominate the
cold component as well. However, this estimate has to be considered as
an upper limit because of potential self-screening effects in the
AGN. Therefore, a significant contribution to the cold component from
more extended star formation cannot be completely excluded.

By looking at the dust SED in W07\footnote{As W07 uses a grey body fit
to the dust SED and B06 only a black body, we base our discussion on
W07.}, the contribution of the 'warm' component to the 200$\mu$m
rest-frame emission is $\sim$70\%. The diameter estimate of 80~pc
based on our data lies within the maximum possible diameter of up to
120-260~pc in W07 for the 'warm' dust component. Even when accounting
for the apparent offset between AGN and dust emission, this may
indicate that the 200~$\mu$m rest-frame emission in \sapm\, is
dominated by the warm dust component although a fractional
contribution of the cold component cannot be entirely
excluded. Following previous assumptions (e.g., W07) and a similar
argumentation as used for Arp~220 \citep[see][]{downes07}, a SB within
a diameter of 80~pc seems unlikely also in \sapm\, leaving the AGN as
the main suspect for the dust heating at 200~$\mu$m: similar to
Arp~220, \sapm\, has a very high intrinsic luminosity (for m=90) of
$\sim$2$\times$10$^{12}$L$_{\odot}$ at 200~$\mu$m (W07) and, hence, an
emission surface brightness of $>$10$^{14}$L$_\odot$~kpc$^{-2}$, which
is 60 times the luminosity of the nearby SB galaxy M82 in a $\sim$4000
times smaller volume \citep[M82's SB disk has a radius of
$\sim$650~pc;][]{garcia00}. Although single super-star clusters (SSCs)
can reach such a high surface brightness of
$\gg$10$^{13}$L$_\odot$~kpc$^{-2}$ \citep[e.g.][]{mart05}, APM08 would
need to harbor $>$1000 SSCs that would fill almost its entire central
region (typical SSC diameters are $\sim$1~pc). This seems to be a
rather unlikely scenario. Even if we consider the uncertainties of our
models, we can clearly constrain the intrinsic radius to be
$\lesssim$150~pc (see Fig.~\ref{fig2}d), unless the dust 'disk' in
\sapm\, had a very small inclination angle ($i$$<$30$^\circ$), which
is, however, not supported by CO observations of W07 who favor
$i$$\approx$70$^\circ$. Also in this more conservative case, the
inferred surface brightness of \sapm\, still exceeds that of any SBs
seen in the present-day universe by more than an order of magnitude,
likely discarding the SB scenario for \sapm.

Based on the flux and the deconvolved size determined from the
submm-observations, the instrinsic brightness temperature at rest of
\sapm\, can be estimated to be T$_{\rm b}$$\approx$30~K. A comparison
to the dust temperature of 220~K of the warm component suggests a dust
opacity of $\tau$$\approx$0.15 (with $\tau$=$-$ln(1-T$_{\rm
b}$/T$_{\rm dust}$)). Assuming the dust temperature of the cold
component of 65~K yields a more conservative upper limit of the
opacity of $\leq$0.6 at 200~$\mu$m rest-frame, translating to an even
smaller upper limit when converted to 1.3~mm. This upper limit is
significantly lower than that found for the AGN in Arp~220 of
$\geq$0.7 at 1.3~mm by \cite{downes07}; the dust emission in Arp~220
has been suggested to be very similar to that of \sapm. This may
indicate that the dust emission in \sapm\, could be either less dense
and/or more clumpy than around the AGN of Arp~220. However, given the
similar (intrinsic) dust masses of $\sim$10$^7$M$_\odot$ and the
similar intrinsic diameters of $\sim$80~pc in both sources, a lower
density in \sapm\, may seem less likely.

\section{Summary \& Conclusion}
\label{con}
We have detected the 200~$\mu$m rest-frame continuum emission in
\sapm\, using the SMA with an angular resolution of
$\sim$0\fasec4. The two (main) lensed images are clearly separated
with a combined flux of $\sim$60$\pm$12~mJy.  Simulations of the
gravitational lensing effect in this system yield a diameter of the
intrinsic (submm) continuum emitting region of $\sim$80~pc and
magnification factor of 90. Our data and simulations seem to be only
consistent with a lensing scenario including a third lensed image of
\sapm\, close to NE if the position of the compact optical/NIR
emission and the position of the extended submm emission are offset
from each other before lensing by $\sim$0\fasec003 ($\equiv$21
pc). Further (sub)mm observations may be beneficial to determine
whether such a positional offset is indeed necessary or more
complicated lens potentials have to be considered. Given our size
estimate, we favor a scenario in which the 200~$\mu$m emission
originates from a warm dust component that is supposed to be mainly
heated by the AGN rather than by a SB.

\acknowledgments The SMA is a joint project between the Smithsonian
Astrophysical Observatory and the Academia Sinica Institute of
Astronomy and Astrophysics and is funded by SAO and ASIAA. We thank
Drs~Downes and Neri for fruitful discussions and the referee for a
thourough review of the paper. We are grateful to the dedicated SMA
staff who make these observations possible.

\clearpage

\begin{deluxetable}{lccccc}
\tabletypesize{\small} 
\tablecaption{Observational Parameters of \apm}
\tablewidth{\hsize} 
\tablehead{
Component & $\Delta\alpha$$^a$ & $\Delta\delta$$^a$ 
& S$^{\rm peak}_{\rm 1mm}$$^b$  & S$^{\rm int}_{\rm 1mm}$$^c$  
& Deconvolved Size \\
& & &  & & maj.$\times$min.,P.A.\\ 
       & (\asec) & (\asec)  & (mJy~beam$^{-1}$) & (mJy) 
& ($''$$\times$$''$,$^\circ$) }
\startdata 
Total$^d$  &  $-$0.10$\pm$0.03 & $-$0.10$\pm$0.03 & 34$\pm$2 & 60$\pm$12
& (0.8$\times$0.7)$\pm$0.1,(52$\pm$5) \\
NE$^e$     &  0.00$\pm$0.04 & 0.00$\pm$0.04 & 21$\pm$2 & 46$\pm$5
& (0.5$\times$0.2)$\pm$0.1,(50$\pm$20) \\
SW$^e$     & $-$0.28$\pm$0.06 & $-$0.26$\pm$0.06 & 15$\pm$2 & 15$\pm$2
& $\leq$0.2 \\
\enddata 

\tablenotetext{a}{\footnotesize\, The offsets are with respect to
$\alpha_{\rm J2000}$=08h31m41.69s and $\delta_{\rm
J2000}$=52d45m17.5s. The positional errors comprise the statistical
errors from the Gaussian fit to the data in the uv-plane and
uncertainties from the calibration. }

\tablenotetext{b}{\footnotesize\, Peak flux. Errors are based on the
statistical noise in the map.}

\tablenotetext{c}{\footnotesize\, Spatially integrated flux. The flux
errors include the statistical uncertainties from the fit and in the
case of the total flux also the calibrational uncertainties, which are
estimated to be $\sim$20\%.}

\tablenotetext{d}{\footnotesize\, Based on the naturally weighted map and an
elliptical gaussian fit. }

\tablenotetext{e}{\footnotesize\, Based on the robustly weighted map and an
elliptical (circular) gaussian fit for NE (SW).}

\label{tab1}
\end{deluxetable}

\begin{deluxetable}{lccccccc}

\tabletypesize{\small} \tablecaption{Parameters of the best-fit
gravitational lens model.}  \tablewidth{0pt} \tablehead{ &
$\theta_{\rm E}$$^b$ & $\epsilon$$^c$ & PA$^d$ & $\theta_{\rm c}$$^e$
& ($\Delta\alpha$,$\Delta\delta$)$^f$ & $\theta_s$$^g$ & $m$$^h$ \\
%& & & & & position$^f$ & & \\ 
& (\asec) & & ($^\circ$) & (\asec) &
(\asec,\asec) & (\asec) } 
\startdata 
%SIS  & 0.25 & - & - & - & ($-$0.18,$-$0.17) & 0.12 & 5 \\ 
%{\bf EP(2)}$^a$ & {\bf 0.29} & {\bf 0.05}  & {\bf 50} & {\bf 0.16} & {\bf ($-$0.178,$-$0.208)} & {\bf 0.04} & {\bf 20} \\
EP(3)$^a$ & 0.31 & 0.008 & -30 & 0.21 & ($-$0.218,$-$0.158) & 0.012 & 90 \\ \enddata

\tablenotetext{a}{\footnotesize \, Elliptical potential with three
lensed images; we allowed the unlensed source position to be variable
between the optical/NIR and our submm data (see text). }

\tablenotetext{b}{Einstein radius $\theta_{\rm E}$; tested parameter
range: $\theta_{\rm E}$=0\fasec1-0\fasec4}

\tablenotetext{c}{Ellipticity $\epsilon$; tested parameter range:
$\epsilon$=0-0.2}

\tablenotetext{d}{Positional angle PA (from N to E): tested range
$-$90 to $+$90$^\circ$}

\tablenotetext{e}{Core radius $\theta_{\rm c}$; tested parameter
range: $\theta_{\rm c}$=0-0.3\asec}

\tablenotetext{f}{The position of the unlensed source is with respect
to the lensed component NE. The lens position is at
($\Delta\alpha$,$\Delta\delta$)=($-$0\farcs210,$-$0\farcs150).}

\tablenotetext{g}{$\theta_s$=intrinsic source diameter; tested range:
0\fasec001-0\fasec3 ($\equiv$0.007-2.1~kpc)}

\tablenotetext{h}{$m$=magnification factor}
\label{tab2}
\end{deluxetable}

\clearpage

\end{document}